\documentstyle[12pt]{article}

\def\8{\infty}
\def\oh{\frac{1}{2}}
\def\ot{\frac{1}{3}}
\def\oq{\frac{1}{4}}
\def\tt{\frac{2}{3}}

\def\d{\partial}
\def\i{\imath\,}
\def\alp{\alpha}
\def\bet{\beta}
\def\gam{\gamma}
\def\lam{\lambda}
\def\eps{\epsilon}

\def\del{\delta}
\def\dal{\partial_{\alpha}}

\def\undertext#1{\vtop{\hbox{#1}\kern 1pt \hrule}}
\def\ra{\rightarrow}
\def\lra{\longrightarrow}

\def\VEV#1{\left\langle #1\right\rangle}

\def\pbyp#1#2{\frac{\partial#1}{\partial#2}}

\def\fbyf#1#2{\frac{\delta#1}{\delta#2}}

\def\br{\\ \nonumber &&}
\def\inv#1{\frac{1}{#1}}
\def\be{\begin{equation}}
\def\ee{\end{equation}}
\def\bea{\begin{eqnarray} &&}
\def\eea{\end{eqnarray}}
\def\ct#1{\cite{#1}}
\def\rf#1{(\ref{#1})}
\def\EXP#1{\exp\left(#1\right)}

\def\1N{$\frac{1}{N}$ expansion}

\def\et{{\cal E}}

\def\NS{Navier-Stokes }
\def\val{v_{\alpha}}
\def\vbe{v_{\beta}}

\def\ral{r_{\alpha}}
\def\rbe{r_{\beta}}

\begin{document}

\begin{titlepage}
{\bf March '93}\hfill	  {\bf PUPT-1383}\\

\begin{center}

{\bf LOOP EQUATION IN TURBULENCE}

\vspace{1.5cm}

{\bf   A.A.~Migdal}

\vspace{1.0cm}

{\it  Physics Department, Princeton University,\\
Jadwin Hall, Princeton, NJ 08544-1000.\\
E-mail: migdal@acm.princeton.edu}

\vspace{1.9cm}
\end{center}

\abstract{
The incompressible fluid dynamics is reformulated as dynamics of
closed loops $C$ in coordinate space. This formulation
allows to derive explicit functional equation for the generating
functional $\Psi[C]$ in inertial range of spatial scales, which
allows the
scaling solutions. The requirement of finite energy dissipation rate
leads then
to  the Kolmogorov index. {\em We find an exact steady solution of
the loop
equation in inertial range of the loop sizes.}  The generating
functional
decreases as $\EXP{-A^{\tt}}$ where $A=\oint_C r \wedge dr$ is the
area inside
the loop. The pdf for the velocity circulation $\Gamma$ is
Lorentzian, with the
width $\bar{\Gamma} \propto A^{\tt} $.
}
\vfill
\end{titlepage}

\tableofcontents

\section{Introduction}

Incompressible fluid dynamics underlies the vast majority of natural
phenomena. It is described by famous Navier-Stokes equation
\begin{equation}
\dot{v}_{\alpha} = \nu \partial_{\beta}^2 v_{\alpha} - v_{\beta}
\partial_{\beta} v_{\alpha} - \partial_{\alpha} p \\;\;
\partial_{\alpha}v_{\alpha} = 0 \label{eq1}
\end{equation}
which is nonlinear, and therefore hard to solve.
This nonlinearity makes life more interesting, though, as it leads to
turbulence. Solving this equation with appropriate initial and
boundary
conditions we expect to obtain the chaotic behavior of velocity
field.

The simplest boundary conditions correspond to infinite space with
vanishing velocity at infinity. We are looking for the translation
invariant probability distribution for velocity field, with infinite
range of the wavelengths. In order to compensate for the energy
dissipation, we
add the usual random force to the \NS equations, with the short
wavelength
support, corresponding to large scale energy pumping.

One way to attempt to describe this probability distribution by the
Hopf generating functional (the angular bracket denote time
averaging, or
ensemble averaging over realizations of the random forces)
\begin{equation}
Z[J] = \left \langle \exp \left( \int d^3 r
J_{\alpha}(r)v_{\alpha}(r)\right)
\right \rangle \label{eq2}
\end{equation}
which is known to satisfy linear functional differential equation
\begin{equation}
\dot{Z} = H\left[J,\frac{\delta}{\delta J} \right] Z
\label{eq3}
\end{equation}
similar to the Schr\"odinger equation for Quantum Field Theory, and
equally hard to solve. Nobody managed to go beyond the Taylor
expansion in source $ J $ , which corresponds to the obvious chain of
equations for the equal time correlation functions of velocity
field in various points in space. The same equations could be
obtained
directly from Navier-Stokes equations, so the Hopf equation looks
useless.

In this work we argue, that one could significantly simplify the Hopf
functional without loosing information about correlation functions.
This simplified functional depends upon the set of 3 periodic
functions of one variable
\begin{equation}
C : r_{\alpha} = C_{\alpha}(\theta)\\;\; 0< \theta< 2\pi
\end{equation}
which set describes the closed loop in coordinate space. The
correlation functions reduce to certain functional derivatives of our
loop functional with respect to $ C(\theta)$ at vanishing loop $ C
\rightarrow 0 $.

The properties of the loop functional at large loop $ C $ also have
physical significance. Like the Wilson loops in Gauge Theory, they
describe the statistics of large scale structures of vorticity field,
which is analogous to the gauge field strength. As we argue in this
paper, the Kolmogorov scaling law corresponds to the loop
functional decreasing as  $ \exp \left(-A^{\frac{2}{3}} \right) $,
where $ A $ is the tensor Area inside the loop. This area law emerges
as a
self-consistent solution of our loop equation in the inertial range
of loops.
By Fourier transformation of the loop functional we obtain the pdf
for the
velocity circulation, which turns out Lorentzian.

In Appendix A we recover the expansion in inverse powers of viscosity
by direct
iterations of the loop equation.

In Appendix  B we study the matrix formulation of the \NS equation,
which may
serve as a basis of the random matrix description of turbulence.

In Appendix  C we study the reduced dynamics, corresponding to the
functional
Fourier transform of the loop functional. We argue, that instead of
3D \NS
equations one can use the 1D equations for the Fourier loop
$P_{\alp}(\theta,t)$.

In Appendix D we discuss the relation between the initial data for
velocity
field and the $P$ field, and we find particular realisation for these
initial
data in terms of the gaussian random variables.

In Appendix E we introduce the generating functional for the scalar
products $
P_{\alp}(\theta)P_{\alp}(\theta') $. The advantage of this functional
over the
original $\Psi[C]$ functional is the smoother continuum limit.

Finally, in Appendix F we discuss the possible numerical
implementations of the
reduced loop dynamics.

These four last Appendixes can be skipped at first reading. They
might be
needed for further development of this approach.

\section{The Loop Calculus}

We suggest to use in turbulence
the following version of the Hopf functional
\begin{equation}
\Psi \left[C \right] = \left \langle \exp
\left(
    \frac{\i }{\nu}  \oint dC_{\alpha}(\theta)
v_{\alpha}\left(C(\theta)\right) \right) \right \rangle \label{eq4}
\end{equation}
which we call the loop functional or the loop field.
It is implied that all angular variable $\theta$ run from $ 0 $ to $
2\pi$
and that all the functions of this variable are $ 2\pi$
periodic.\footnote{This
parametrization of the loop is a matter of convention, as the loop
functional
is parametric invariant.} The viscosity $ \nu $ was inserted in
denominator in
exponential, as the only parameter of proper dimension. As we shall
see below,
it plays the role, similar to
the Planck's constant in Quantum mechanics, the turbulence
corresponding to the WKB limit $ \nu \rightarrow 0 $. \footnote{One
could also
insert any numerical parameter in exponential, but this factor could
be
eliminated by space- and/or time rescaling.}

As for the imaginary unit $\i$, there are two reasons to insert it in
the
exponential. First, it makes the motion compact: the phase factor
goes around
the unit circle, when the velocity field fluctuates. So, at large
times one may
expect the ergodicity, with well defined average functional bounded
by $1$ by
absolute value. Second, with this factor of $\i$, the irreversibility
of the
problem is manifest. The time reversal corresponds to the complex
conjugation
of $\Psi$, so that imaginary part of the asymptotic value of $\Psi$
at $t \ra
\8$ measures the effects of dissipation.

The loop orientation reversal $ C(\theta) \ra C(2\pi - \theta) $ also
leads to
the complex conjugation,  so it is equivalent to the time reversal.
This
symmetry implies, that any  correlator of odd/even number of
velocities should
be integrated odd/even number of times over the loop, and it must
enter with an
imaginary/real factor. Later, we shall use this property in the area
law.

We shall often use the field theory notations for the loop integrals,
\be
\Psi \left[C \right] = \left \langle \exp
\left(
    \frac{\i }{\nu}  \oint_C dr_{\alpha}v_{\alpha}
 \right) \right \rangle \label{eq4'}
\ee
This loop integral can be reduced to the surface
integral of vorticity field
\be
\omega_{\mu\nu} = \d_{\mu}v_{\nu}-\d_{\nu}v_{\mu}
\ee
by the Stokes theorem
\begin{equation}
\Gamma_C[v] \equiv \oint_C dr_{\alpha}v_{\alpha}= \int_{S}  d
\sigma_{\mu\nu} \omega_{\mu\nu} \\;\; \partial S = C
\end{equation}

This is the well-known velocity circulation, which measures the net
strength of the vortex lines, passing through the loop $ C $.  Would
we fix initial loop $ C $ and let it move with the flow, the loop
field would be conserved by the Euler equation, so that only the
viscosity effects would be responsible for its time evolution.
However, this is not what we are trying to do.
We take the Euler rather than Lagrange dynamics, so that the loop is
fixed in
space, and hence $\Psi$ is time dependent already in the Euler
equations. The
difference between Euler and \NS equations is the time
irreversibility, which
leads to complex average $\Psi$ in \NS dynamics.

It is implied that this field $\Psi\left[C\right]$ is invariant under
translations of the loop $ C(\theta) \rightarrow C(\theta)+ const $.
The
asymptotic behavior at large time  with proper random forcing reaches
certain
fixed point, governed by the translation- and scale invariant
equations, which
we derive in this paper.

The general Hopf functional (\ref{eq2}) reduces for the loop field
for
the following  imaginary singular source
\begin{equation}
J_{\alpha}(r) =   \frac{\i }{ \nu} \oint_C dr'_{\alpha} \delta^3
\left(r'-r \right) \label{eq8}
\end{equation}

The $\Psi$  functional involves connected correlation functions of
the powers
of circulation at equal times.
\be
\Psi[C] =
\EXP{\sum_{n=2}^{\8}\frac{\i^{n-1}}{n!\,\nu^{n-1}}\,\VEV{\VEV{
\Gamma_C^n[v]}}}
\ee
This expansion goes in powers of effective Reynolds number, so it
diverges in
turbulent region. There, the opposite WKB approximation will be used.

Let us come back to the general case of the arbitrary Reynolds
number. What
could be the use of such restricted Hopf functional?
At first glance it seems that we lost most of information, described
by the Hopf functional, as the general Hopf source $J$ depends upon 3
variables $ x,y,z $ whereas the loop $C$ depends of only one
parameter $ \theta $.
Still, this information can be recovered by taking the loops of the
singular shape, such as two infinitesimal loops $R_1, R_2 $,
connected by a couple of wires
\input epsf \centerline{ \epsfbox{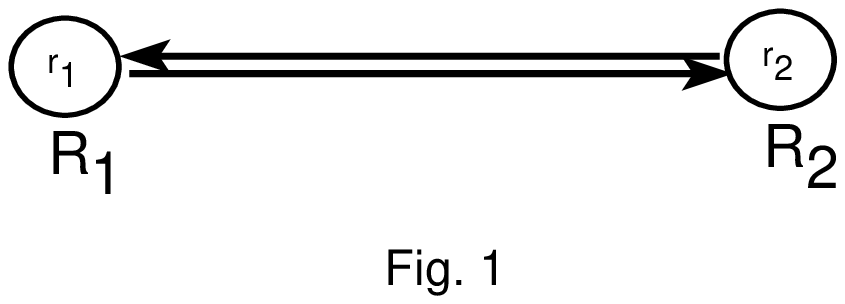}}

The loop field in this case reduces to
\begin{equation}
\Psi \left[C \right] \rightarrow \left \langle  \exp
\left(
    \frac{\i}{ 2\nu} \Sigma_{\mu\nu}^{R_1}\omega_{\mu\nu}(r_1)
+\frac{\i}{ 2\nu}   \Sigma_{\mu\nu}^{R_2 } \omega_{\mu\nu}(r_2)
\right) \right \rangle
\end{equation}
where
\begin{equation}
\Sigma_{\mu\nu}^R = \oint_R d r_{\nu}r_{\mu}
\end{equation}
is the tensor area inside the loop $R$. Taking functional derivatives
with respect to the shape of $R_1$ and $R_2$ prior to shrinking them
to points,
we can bring down the product of vorticities at
$r_1$ and $r_2$.  Namely, the variations yield
\be
\delta\Sigma_{\mu\nu}^R=  \oint _R\left(d r_{\nu}\delta r_{\mu}+
r_{\mu}d
\delta r_{\nu} \right) =  \oint_R \left( d r_{\nu}\delta r_{\mu} -d
r_{\mu}\delta r_{\nu} \right)
\ee
where integration by parts was used in the second term.

One may introduce the area derivative $\fbyf{}{\sigma_{\mu\nu}(r)}$,
which
brings down the vorticity at the given point $ r $ at the loop.
\begin{equation}
-\nu^2 \frac{\delta^2 \Psi \left[C \right]}
{\delta \sigma_{\mu\nu}(r_1)\delta \sigma_{\lambda \rho}(r_2)}
\ra \left \langle \omega_{\mu\nu}(r_1)
\omega_{\lambda \rho}(r_2) \right \rangle
\end{equation}

The careful definition of these area derivatives are or paramount
importance to us. The corresponding loop calculus was developed
in\cite{Mig83} in the context of the gauge theory. Here we rephrase
and further refine the definitions and relations established in that
work.

The basic element of the loop calculus is what we suggest to call the
spike
derivative, namely the operator which adds the infinitesimal $
\Lambda $ shaped spike to the loop
\begin{equation}
D_{\alpha}(\theta,\epsilon) = \int_{\theta}^{\theta+2\epsilon}d \phi
\left(
1-\frac{\left|\theta +\epsilon - \phi\right|}{\epsilon }
\right)
    \frac{\delta}{\delta C_{\alpha}(\phi)}
\end{equation}
The finite spike operator
\begin{equation}
\Lambda(r,\theta,\epsilon) =
\exp \left( r_{\alpha}  D_{\alpha}(\theta,\epsilon) \right)
\end{equation}
adds the spike of the height $r$. This is the straight line from $
C(\theta) $ to $ C(\theta + \epsilon) + r$, followed by another
straight line from $ C(\theta+\epsilon)+r $ to $ C(\theta+2
\epsilon)$,
\input epsf \centerline{ \epsfbox{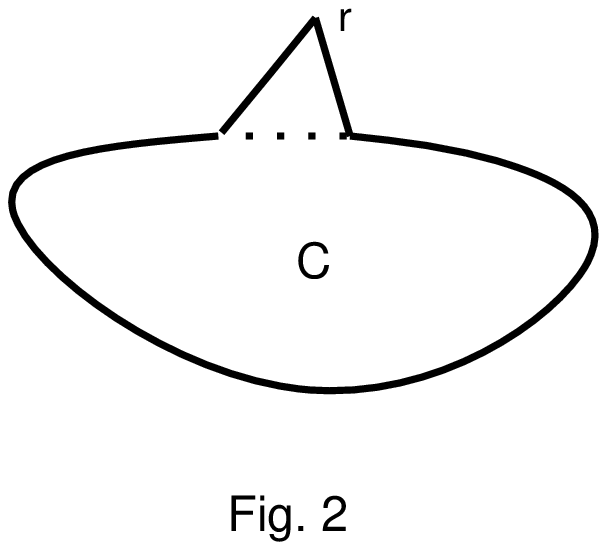}}
Note, that the loop remains
closed, and the slopes remain finite, only the second derivatives
diverge. The continuity and closure of the loop eliminates the
potential part of velocity; as we shall see below,
this is necessary to obtain the loop equation.

In the limit $ \epsilon \rightarrow 0 $ these spikes are invisible,
at
least for the smooth vorticity field, as one can see from the Stokes
theorem (the area inside the spike goes to zero as $ \epsilon $).
However, taking certain derivatives prior to the limit $ \epsilon
\rightarrow 0 $ we can obtain the finite contribution.

Let us consider the operator
\begin{equation}
\Pi \left(r,r',\theta ,\epsilon \right) =
\Lambda  \left(r, \theta,\frac{1}{2} \epsilon \right) \Lambda
\left(r',\theta,\epsilon \right)
\end{equation}
By construction it inserts the smaller spike on top of a bigger one,
in such a way, that a polygon appears
\input epsf \centerline{ \epsfbox{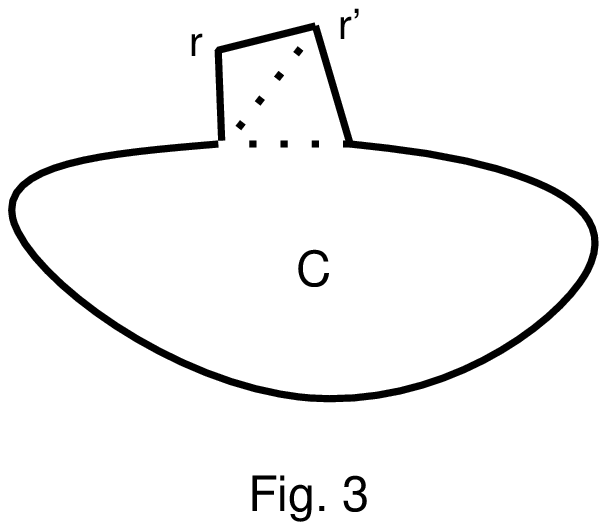}}
Taking the derivatives with respect to the  vertices of
this polygon $ r, r' $ , setting $r=r'=0$ and
antisymmetrising, we find the tensor operator
\begin{equation}
\Omega_{\alpha\beta}(\theta,\epsilon) =
-\i \nu  D_{\alpha}\left(\theta,\frac{1}{2} \epsilon \right)
D_{\beta}\left(\theta,\epsilon \right) - \{\alpha
\leftrightarrow\beta\}
\label{OM}
\end{equation}
which brings down the vorticity, when applied to the loop field
\begin{equation}
\Omega_{\alpha\beta}(\theta,\epsilon) \Psi \left[C \right]
\stackrel{\epsilon \rightarrow 0}{\longrightarrow}
\omega_{\alpha\beta}\left(C(\theta)\right)\Psi \left[C \right]
\label{eqom}
\end{equation}

The quick  way to check these formulas is to use formal functional
derivatives
\begin{equation}
\frac{\delta \Psi \left[C \right]}{\delta C_{\alpha}(\theta)} =
C'_{\beta}(\theta) \fbyf{\Psi \left[C
\right]}{\sigma_{\alp\bet}\left(C(\theta)\right)}
\end{equation}
Taking one more functional derivative derivative we find the term
with
vorticity times first derivative of the $ \delta $ function, coming
from the
variation of $ C'(\theta) $
\be
\frac{\delta^2 \Psi [C ]}{\delta C_{\alp}(\theta) \delta
C_{\bet}(\theta')} =
\del'(\theta-\theta')\fbyf{\Psi \left[C
\right]}{\sigma_{\alp\bet}\left(C(\theta)\right)} +
C'_{\gam}(\theta) C'_{\lam}(\theta')\frac{\delta^2\Psi \left[C
\right]}{\delta
\sigma_{\alp\gam}\left(C(\theta)\right) \delta
\sigma_{\bet\lam}\left(C(\theta')\right)}
\ee
This term is the only one, which survives the limit $ \epsilon
\rightarrow 0 $ in our relation (\ref{eqom}).

So, the area derivative can be defined from the antisymmetric tensor
part of
the second functional derivative as the coefficient in front of $
\delta'(\theta-\theta') $ .  Still, it has all the properties of the
first
functional derivative, as it can also be defined from the above first
variation.
The advantage of dealing with spikes is the control over the limit
$\eps \ra 0$
, which might be quite singular in applications.

So far we managed to insert the vorticity at the loop $ C $ by
variations of the loop field. Later we shall need the vorticity off
the loop, in arbitrary point in space. This can be achieved by the
following combination of the spike operators
\begin{equation}
\Lambda \left(r,\theta,\epsilon \right) \Pi
\left(r_1,r_2,\theta+\epsilon,\delta \right) \\;\; \delta \ll
\epsilon
\end{equation}
This operator inserts the $ \Pi $ shaped little loop at the top of
the
bigger spike, in other words, this little loop is translated by a
distance $r$ by the big spike.

Taking derivatives, we find the operator of finite translation of the
vorticity
\begin{equation}
\Lambda \left(r,\theta,\epsilon \right)
\Omega_{\alpha\beta}(\theta+ \epsilon ,\delta)
\end{equation}
and the corresponding infinitesimal translation operator
\begin{equation}
D_{\mu}(\theta,\epsilon)\Omega_{\alpha\beta}(\theta+ \epsilon
,\delta)
\end{equation}
which inserts $ \partial_{\mu} \omega_{\alpha \beta} \left( C(\theta)
\right) $ when applied to the loop field.

Coming back to the correlation function, we are going now to
construct  the
operator, which would insert two vorticities separated by a distance.
Let us
note that the global $ \Lambda $ spike
\begin{equation}
\Lambda \left(r,0,\pi \right) = \exp
\left(
    r_{\alpha}\int_{0}^{2\pi}d
\phi  \left(1- \frac{ \left|\phi-\pi \right|}{\pi} \right)
\frac{\delta}{\delta C_{\alpha}(\phi)}\right)
\end{equation}
when applied to a  shrunk loop $ C(\phi) = 0 $ does nothing but
the backtracking from $0$ to $r$
\input epsf \centerline{ \epsfbox{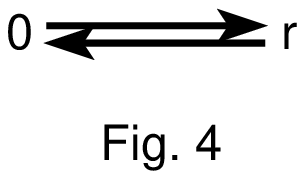}}
This means that the operator
\begin{equation}
\Omega_{\alpha\beta}(0 ,\delta)\Omega_{\lambda \rho}(\pi ,\delta)
\Lambda \left(r,0,\pi \right)
\end{equation}
when applied to the loop field for a shrunk loop yields the vorticity
correlation function
\begin{equation}
\Omega_{\alpha\beta}(0 ,\delta)\Omega_{\lambda \rho}(\pi ,\delta)
\Lambda \left(r,0,\pi \right) \Psi [0] = \left \langle \omega_{\alpha
\beta}(0) \omega_{\lambda \rho}(r) \right \rangle
\end{equation}

The higher correlation functions of vorticities could be constructed
in a
similar fashion, using the spike operators. As for the velocity, one
should
solve the Poisson equation
\begin{equation}
\partial_{\mu}^2 v_{\alpha}(r) = \partial_{\beta} \omega_{\beta
\alpha}(r)
\end{equation}
with the proper boundary conditions , say, $ v=0 $ at infinity.
Formally,
\begin{equation}
v_{\alpha}(r) =
\frac{1}{\partial_{\mu}^{2}}\partial_{\beta} \omega_{\beta \alpha}(r)
\end{equation}

This suggests the following formal definition of  the velocity
operator
\begin{equation}
V_{\alpha}(\theta,\epsilon,\delta) =
\frac{1}{D_{\mu}^2(\theta,\epsilon)}
D_{\beta}(\theta,\epsilon) \Omega_{\beta \alpha}(\theta,\delta)\\;\;
\delta \ll \epsilon
\label{VOM}
\end{equation}
\begin{equation}
V_{\alpha}(\theta,\epsilon,\delta)\Psi[C] \stackrel{\delta,\epsilon
\rightarrow 0}{\longrightarrow} v_{\alpha} \left(C(\theta) \right)
\Psi[C]
\end{equation}

Another version of this formula is the following integral
\begin{equation}
V_{\alpha}(\theta,\epsilon,\delta)= \int d^3  \rho
\frac{\rho_{\beta}}{4 \pi |\rho|^3}\Lambda \left(\rho,\theta,\epsilon
\right)
\Omega_{\alpha\beta}(\theta+ \epsilon ,\delta)
\end{equation}
where the $ \Lambda $ operator shifts the  $ \Omega $ by a distance $
\rho $ off the original loop at the point $ r = C(\theta + \epsilon)
$
\input epsf \centerline{ \epsfbox{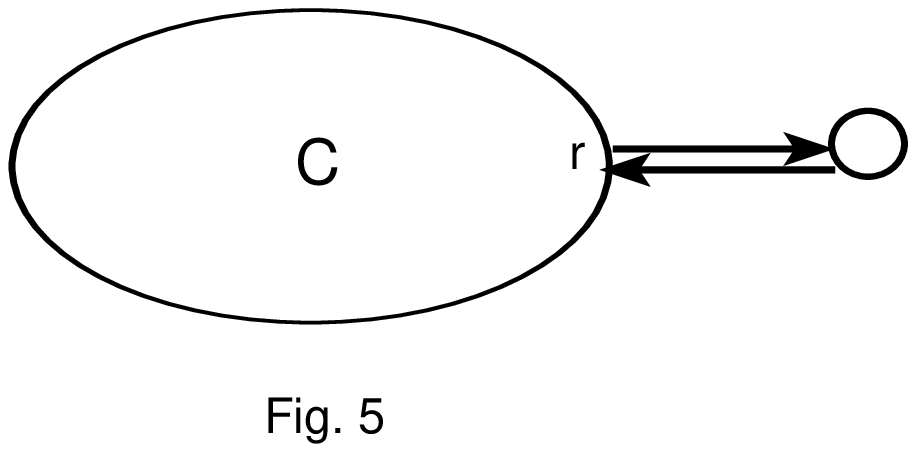}}

\section{Loop Equation}

Let us now derive exact equation for the loop functional.
Taking the time derivative of the original definition, and using the
Navier-Stokes equation we get in front of exponential
\begin{equation}
\oint_C d r_{\alpha}  \frac{\i}{ \nu}
\left(
    \nu \partial_{\beta}^2 v_{\alpha} - v_{\beta}
\partial_{\beta} v_{\alpha} - \partial_{\alpha} p \right)
\end{equation}
The term with the pressure gradient yields zero after integration
over
the closed loop, and the velocity gradients in the first two terms
could be expressed in terms of vorticity up to irrelevant gradient
terms, so that we find
\begin{equation}
\oint_C d r_{\alpha}  \frac{\i}{ \nu}
\left(
    \nu \partial_{\beta} \omega_{\beta \alpha} - v_{\beta}
\omega_{\beta \alpha}
\right) \label{Orig}
\end{equation}

Replacing the vorticity and velocity  by the operators discussed in
the
previous Section we find the following loop equation (in explicit
notations)
\bea
-\i\dot{\Psi}[C] = \br
\oint d C_{\alpha}(\theta)
\left(
    D_{\beta}(\theta,\epsilon) \Omega_{\beta \alpha}(\theta,\epsilon)
+
\frac{1}{ \nu}
\int d^3  \rho
\frac{\rho_{\gam}}{4 \pi |\rho|^3}\Lambda \left(\rho,\theta,\epsilon
\right)
\Omega_{\gam\bet}(\theta+ \epsilon ,\delta)\Omega_{\beta
\alpha}(\theta,\delta)
\right) \Psi[C]
\label{PsiC}
\eea

The more compact form of this equation, using the notations of
\cite{Mig83}, reads
\bea
\i\,\nu\dot{\Psi}[C] = {\cal H}_C\Psi \br
{\cal H}_C \equiv \nu^2\oint_{C} dr_{\alpha}
\left(
    \i\partial_{\beta} \frac{\delta }{\delta \sigma_{\beta
\alpha}(r)}+
\int d^3 r'\frac{r'_{\gamma}-r_{\gam}}{4 \pi |r-r'|^3}
\frac{\delta^2}{\delta \sigma_{\beta \alpha}(r)
\delta \sigma_{\beta \gamma}(r')}
\right)
\label{OLD}
\eea
Now we observe that viscosity $ \nu $ appears in front of time  and
spatial
derivatives, like the Planck constant $\hbar$ in Quantum mechanics.
Our loop
hamiltonian ${\cal H}_C$ is not hermitean, due to dissipation. It
contains the
second loop derivatives, so it represents a (nonlocal!) kinetic term
in loop
space.

So far, we considered so called decaying turbulence, without external
energy
source. The  energy
\be
E = \int d^3 r \oh \, \val^2
\ee
would eventually all dissipate, so that the fluid would stop. In this
case the
loop wave function $\Psi$ would asymptotically approach $1$
\be
\Psi[C] \stackrel{t \ra \8}\lra 1
\ee

In order to reach the steady state, we  add to the right side of the
\NS
equation the usual gaussian random forces $f_{\alp}(r,t)$ with the
space
dependent correlation function
\be
\VEV{f_{\alp}(r,t)f_{\bet}(r',t')} =
\delta_{\alp\bet}\delta(t-t')F(r-r')
\ee
concentrated at at small wavelengths, i.e. slowly varying with
$r-r'$.

Using the identity
\be
\VEV{f_{\alp}(r,t) \Phi[v(.)]} = \int d^3 r' F(r-r')
\fbyf{\Phi[v(.)]}{\val(r')}
\ee
which is valid for arbitrary functional $\Phi$  we find the following
imaginary potential term in the loop hamiltonian
\be
\delta{\cal H}_C \equiv \i\,U[C]= \frac{\i}{\nu}\,\oint_{C}
dr_{\alpha}\oint_{C} dr'_{\alpha} F(r-r')
\ee

Note, that  orientation reversal together with complex conjugation
changes the
sign of the loop hamiltonian, as it should. The potential part
involves two
loop integrations times imaginary constant. The first term in the
kinetic part
has one loop integration, one loop derivative times imaginary
constant. The
second kinetic term has one loop integration, two loop derivatives
and real
constant. The left side of the loop equation has no loop
integrations, no loop
derivatives, but has a factor of $\i$.

The relation between the potential and kinetic parts of the loop
hamiltonian
depends of viscosity, or, better to say, it depends upon the Reynolds
number,
which is the ratio of the typical circulation to viscosity. In the
viscous
limit, when the Reynolds number is small, the loop wave function is
close to
$1$. The perturbation expansion in $ \inv{\nu}$ goes in powers of the
potential, in the same way, as in Quantum mechanics. The second
(nonlocal) term
in kinetic part of the hamiltonian also serves as a small
perturbation (it
corresponds to nonlinear term in the \NS equation).
The first term of this perturbation expansion is just
\be
\Psi[C] \ra 1 - \int \frac{d^3
k}{(2\pi)^3}\frac{\tilde{F}(k)}{2\nu^3\,k^2}
\left|\oint_C d \ral e^{\i k r}\right|^2
\ee
with $\tilde{F}(k)$ being the Fourier transform of $F(r)$.
This term is real, as it corresponds to the two-velocity correlation.
The next
term comes from the triple correlation of velocity, and this term is
purely
imaginary, so that the dissipation shows up.

This expansion can be derived  by direct iterations in the loop space
as in
\cite{Mig83}, inverting the operator in the local part of the kinetic
term in
the hamiltonian.  This expansion is discussed in Appendix A. The
results agree
with the straightforward iterations of  the \NS equations in powers
of the
random force, starting from zero velocity.

So, we have the familiar situation, like in QCD, where the
perturbation theory
breaks because of the infrared divergencies. For arbitrarily small
force, in a
large system, the region of small $k$ would yield large contribution
to the
terms of the perturbation expansion. Therefore, one should take the
opposite
WKB limit $\nu \ra 0$.

In this limit, the wave function should behave as the usual WKB wave
function,
i.e. as an exponential
\be
\Psi[C] \ra \EXP{\frac{\i\,S[C]}{\nu}}
\ee
The effective loop Action $S[C]$ satisfies the loop space
Hamilton-Jacobi
equation
\be
\dot{S}[C] =-\i U[C] + \oint_{C} dr_{\alpha}
\int d^3 r'\frac{r'_{\gamma}-r_{\gam}}{4 \pi |r-r'|^3}
\frac{\delta S}{\delta \sigma_{\beta \alpha}(r)}
\frac{\delta S}{\delta \sigma_{\beta \gamma}(r')}
\label{SC}
\ee
The imaginary part of $S[C]$ comes from imaginary potential $U[C]$,
which
distinguishes our theory from the reversible Quantum mechanics. The
sign of
$\Im S$ must be positive definite,  since $ |\Psi| <1$. As for  the
real part
of $S[C]$, it changes the sign under the loop orientation reversal
$C(\theta)
\ra C(2\pi-\theta) $.

At finite viscosity there would be an additional term
\be
-\nu\oint_{C} dr_{\alpha}\partial_{\beta} \frac{\delta S[C]}{\delta
\sigma_{\beta\alpha}(r)}
-\i\nu \oint_{C} dr_{\alpha}
\int d^3 r' \frac{r'_{\gamma}-r_{\gam}}{4 \pi |r-r'|^3}
\frac{\delta^2 S[C]}{\delta \sigma_{\beta \alpha}(r) \delta
\sigma_{\beta
\gamma}(r')}
\ee
on the right of \rf{SC}.  As for the term
\be
-\oint_{C} dr_{\alpha}
    \i\left(\partial_{\beta}S[C]\right) \frac{\delta S[C]}{\delta
\sigma_{\beta
\alpha}(r)}
\ee
which formally arises in the loop equation, this term vanishes, since
$\partial_{\beta}S[C]=0$. This operator inserts backtracking at some
point at
the loop without first applying the loop derivative at this point. As
it was
discussed in the previous Section, such backtracking  does not change
the loop
functional. This issue was discussed at length in \ct{Mig83}, where
the
Leibnitz rule for the operator $ \dal \fbyf{}{\sigma_{\bet\gam}} $
was
established
\be
\dal \fbyf{f(g[C])}{\sigma_{\bet\gam}(r)} = f'(g[C])\dal
\fbyf{g[C]}{\sigma_{\bet\gam}(r)}
\ee
In other words, this operator acts as a first order derivative on the
loop
functional with finite area derivative (so called Stokes type
functional).
Then, the above term does not appear.

The Action functional $ S[C] $ describes the distribution of the
large scale
vorticity structures, and hence it should not depend of viscosity. In
terms of
the above connected correlation functions of the circulation this
corresponds
to the limit, when the effective Reynolds number
$\frac{\Gamma_C[v]}{\nu}$ goes
to infinity, but the sum of the divergent series tends to the finite
limit.
According to the standard picture of turbulence, the large scale
vorticity
structures depend upon the energy pumping, rather than the energy
dissipation.

This , of course implies, that  both time $ t $ and the loop
size\footnote{As a
measure of the loop size one may take the square root of the minimal
area
inside the
loop.} $ |C| $ should be greater then the viscous scales
\begin{equation}
t \gg t_0 = \nu^{\frac{1}{2}}{\cal E}^{-\frac{1}{2}} \\;\;
|C| \gg r_0 = \nu^{\frac{3}{4}} {\cal E}^{-\frac{1}{4}}
\end{equation}
where $ {\cal E } $ is the energy dissipation rate.

It is defined from the energy balance equation
\be
0 = \d_t\VEV{\oh\,\val^2}= \nu \VEV{\val\d^2 \val}
+\VEV{f_{\alp}\val}
\ee
which can be transformed to
\be
 \oq \nu \VEV{\omega_{\alp\bet}^2} = 3 F(0)
\ee
The left side represents the energy, dissipated at small scale due to
viscosity, and the right side - the energy pumped in from the large
scales due
to the random forces. Their common value is $\et$.

We see, that constant $F(r-r')$, i.e., $\tilde{F}(k)\propto
\delta(k)$ is
sufficient to  provide the necessary energy pumping. However, such
forcing does
not produce vorticity, which we readily see in our equation. The
contribution
from this constant part to the potential in our loop equation drops
out (this
is a  closed loop integral of total derivative). This is important,
because
this term would have the wrong order of magnitude in the turbulent
limit - it
would grow as the Reynolds number.

Dropping this term, we arrive at remarkably simple and universal
functional
equation
\be
\dot{S}[C] =  \oint_{C} dr_{\alpha}
\int d^3 r'\frac{r'_{\gamma}-r_{\gam}}{4 \pi |r-r'|^3}
\frac{\delta S}{\delta \sigma_{\beta \alpha}(r)}
\frac{\delta S}{\delta \sigma_{\beta \gamma}(r')}
\label{KIN}
\ee
The stationary solution of this equation describes the steady
distribution of
the circulation in the strong turbulence. Note, that the  stationary
solutions
come in pairs $ \pm S$. The sign should be chosen so, that $ \Im S >
0 $, to
provide the inequality $ |\Psi| <1$.

\section{Scaling law}

The `Hamilton-Jacobi' equation  without the potential term
(\ref{KIN}) allows
the family of the scaling solutions
\begin{equation}
S[C] = t^{2 \kappa -1}\phi \left[\frac{C}{t^{\kappa}} \right]
\end{equation}
with arbitrary index $ \kappa $. The scaling function satisfies the
equation
\begin{equation}
(2 \kappa -1 ) \phi[C]  - \kappa \oint_{C} dr_{\alpha}
\frac{\delta \phi[C]}{\delta \sigma_{\beta \alpha}(r)}r_{\beta} =
\oint_{C} dr_{\alpha}
\int d^3 r'\frac{r'_{\gamma}-r_{\gam}}{4 \pi |r-r'|^3}
\frac{\delta \phi[C]}{\delta \sigma_{\beta \alpha}(r)}
\frac{\delta \phi[C]}{\delta \sigma_{\beta \gamma}(r')}
\end{equation}
The left side here was computed, using the chain rule differentiation
of
functional.

Asymptotically, at large time, we expect the fixed point, which is
the
homogeneous functional
\begin{equation}
S_{\infty}[C] = |C|^{2- \frac{1}{\kappa}} f \left[\frac{C}{|C|}
\right]
\end{equation}
zeroing the right side of our `kinetic' functional equation
\begin{equation}
0=\oint_{C} dr_{\alpha}
\int d^3 r'\frac{r'_{\gamma}-r_{\gam}}{4 \pi |r-r'|^3}
\frac{\delta S_{\infty}[C]}{\delta \sigma_{\beta \alpha}(r)}
\frac{\delta S_{\infty}[C]}{\delta \sigma_{\beta \gamma}(r')}
\end{equation}

The Kolmogorov scaling \cite{Kolm41} would correspond to
\begin{equation}
\kappa = \frac{3}{2}
\end{equation}
in which case one can express the $ S $ functional in terms of $ {
\cal E } $
\begin{equation}
S[C] = {\cal E} t^2 \phi \left[\frac{C}{\sqrt{{\cal E}t^3}} \right]
\end{equation}

One can easily rephrase the Kolmogorov arguments in the loop space.
The relation between the energy dissipation rate and the velocity
correlator reads
\begin{equation}
{\cal E } = \left \langle v_{\alpha}(r_0) v_{\beta}(0)
\partial_{\beta}
v_{\alpha}(0) \right \rangle
\end{equation}
where the point splitting at the viscous scale $r_0$ is introduced.
Such
splitting is necessary to avoid the viscosity effects; without the
splitting the average would formally reduce to the total derivative
and vanish.

Instead of the point splitting one may introduce the finite loop of
the viscous scale $ |C| \sim r_0 $, and compute this correlator in
presence of
such
loop. This reduces to the WKB estimates
\begin{equation}
\omega_{\alpha \beta}(r) \rightarrow
\frac{\delta S[C]}{\delta \sigma_{\alpha \beta}(r)} \\;\;
v_{\alpha}(r) = \int d^3 r'\frac{r'_{\gamma}-r_{\gam}}{4 \pi
|r-r'|^3}
\omega_{\alpha \gamma}(r')
\end{equation}

Using the generic scaling law for $ S $ we find
\begin{equation}
\omega \sim r_0^{- \frac{1}{\kappa}}\\;\;
v \sim r_0^{1-  \frac{1}{ \kappa}}\\;\;
{\cal E} \sim r_0^{2 - \frac{3}{\kappa}}
\end{equation}

We see, that the energy dissipation rate would stay finite in the
limit of the
vanishing viscous scale only for the Kolmogorov value of the index.
This
argument looks rather cheap, but I think it is basically
correct. The constant value of the energy dissipation rate in the
limit
of vanishing viscosity arises as the quantum anomaly in the field
theory, through the finite limit of the point splitting in the
correspondent energy current.\footnote{I am grateful to A.~Polyakov
and
E.~Siggia for inspiring comments on this subject.}

There is another version of this argument, which I like better. The
dynamics of
 Euler fluid in infinite system would not exist, for the
non-Kolmogorov
scaling. The extra powers of loop size would have to enter with the
size $L$ of
the whole system, like $\left(\frac{|C|}{L}\right)^{\eps} $. So, in
the regime
with finite energy pumping rate $\et$ the infinite Euler system can
exist only
for the Kolmogorov index. This must be the essence of the original
Kolmogorov
reasoning \ct{Kolm41}.

The problem is that nobody proved that such limit exists, though.
Within the
usual framework, based on the velocity correlation functions, one has
to prove,
that the infrared divergencies, caused by the sweep, all cancel for
the
observables. Within our framework these problems disappear, as we
shall see
later.

As for the correlation functions in inertial range, unfortunately
those cannot be computed in the WKB approximation, since they involve
the
contour shrinking to a double line, with vanishing area inside.
Still, most of
the physics can be understood in loop terms, without these
correlation
functions. The large scale behavior of the loop functional reflects
the
statistics of the large vorticity structures, encircled by the loop.

\section{Area law}

The Wilson loop in QCD decreases as exponential of the minimal area,
encircled
by the loop, leading to the quark confinement. What is the similar
asymptotic
law in turbulence? The physical mechanisms leading to the area law in
QCD are
absent here. Moreover, there is no guarantee, that $\Psi[C]$ always
decreases
with the size of the loop.

This makes it possible to look for the simple Anzatz, which was not
acceptable
in QCD, namely
\be
S[C] = s\left(\Sigma_{\mu\nu}^C\right)
\ee
where
\be
\Sigma_{\mu\nu}^C= \oint_C r_{\mu} d r_{\nu}
\ee
is the tensor area encircled by the loop $C$. The difference between
this area
and the scalar area is the positivity property. The scalar area
vanishes only
for the loop which can be contracted to a point by removal of all the
backtracking. As for the tensor area, it vanishes, for example, for
the $8$
shaped loop, with opposite orientation of petals.

Thus, there are some large contours with vanishing tensor area, for
which there
would be no decrease of the $\Psi$ functional.
In QCD the Wilson loops must always decrease at large distances, due
to the
finite mass gap. Here, the large scale correlations are known to
exist, and
play the central role in the turbulent flow. So, I  see no reasons to
reject
the tensor area Anzatz.

This Anzatz in QCD not only was unphysical, it failed to reproduce
the correct
short-distance singularities in the loop equation. In turbulence,
there are no
such singularities. Instead, there are the  large-distance
singularities, which
all should cancel in the loop equation.

It turns out, that  for this Anzatz the (turbulent limit of the) loop
equation
is satisfied automatically, without any further restrictions.
Let us verify this important property. The first area derivative
yields
\be
\omega_{\mu\nu}^C(r)=\fbyf{S}{\sigma_{\mu\nu}(r)} = 2\pbyp{s}{
\Sigma_{\mu\nu}^C}
\ee
The factor of $2$ comes from the second term in the variation
\be
\fbyf{\Sigma^C_{\alp\bet}}{\sigma_{\mu\nu}(r)}=
\del_{\alp\mu}\del_{\bet\nu}-\del_{\alp\nu}\del_{\bet\mu}
\ee
Note, that the right side does not depend on $r$. Moreover, you can
shift $r$
aside from the base loop $C$, with proper wires inserted. The area
derivative
would not change, as the contribution of wires drops.

This implies, that the corresponding vorticity $\omega_{\mu\nu}^C(r)
$ is space
independent, it only depends upon the loop itself. The velocity can
be
reconstructed from vorticity up to irrelevant potential terms
\be
\vbe^C(r) = \oh\,\ral\,\omega_{\alp\bet}^C
\ee
This can be formally obtained from the above integral representation
\be
\vbe^C(r) =\int d^3 r'\frac{\ral-r'_{\alp}}{4 \pi
|r-r'|^3}\omega_{\alp\bet}^C
\label{INTG}
\ee
as a residue from the infinite sphere $ R = |r'| \ra \8$. One may
insert the
regularizing factor $ |r'|^{-\epsilon}$ in $\omega$, compute the
convolution
integral in Fourier space and check that in the limit $ \epsilon  \ra
0^+$ the
above linear term arises. So, one can use the above form of the loop
equation,
with the analytic regularization prescription.

Now, the $v\,\omega$ term in the loop equation reads
\be
\oint _C d r_{\gam} \,\vbe^C(r)\,\omega_{\bet\gam}^C \propto
\Sigma_{\gam\alp}^C\,\omega_{\alp\bet}^C\,\omega_{\bet\gam}^C
\ee
This tensor trace vanishes, because the first tensor is
antisymmetric, and the
product of the last two antisymmetric tensors is symmetric with
respect to
$\alp\gam$.

So, the positive and negative terms cancel each other in our loop
equation,
like the "income" and "outcome" terms in the usual kinetic equation.
We see,
that there is an equilibrium  in our loop space kinetics.

{}From the point of view of the notorious infrared divergencies in
turbulence,
the above calculation explicitly demonstrates how they cancel. By
naive
dimensional counting these terms were linearly divergent. The space
isotropy
lowered this to logarithmic divergency in \rf{INTG}, which reduced to
finite
terms at closer inspection. Then, the explicit form of these terms
was such,
that they all cancelled.

This cancellation originates from the angular momentum conservation
in fluid
mechanics.  The large loop $C$  creates the macroscopic eddy with
constant
vorticity  $\omega_{\alp\bet}^C$ and linear velocity $ v^C(r) \propto
r$. This
is a well known static solution of the \NS equation. The eddy is
conserved due
to the angular momentum conservation.The only nontrivial thing  is
the
functional dependence of the eddy vorticity upon the shape and size
of the loop
$C$.  This is a function of the tensor area $\Sigma_{\mu\nu}^C$,
rather than a
general functional of the loop.

Combining this Anzatz with the space isotropy and the Kolmogorov
scaling law,
we arrive at the turbulent area law
\be
\Psi[C] \propto \EXP{-
B\,\left(\frac{\et}{\nu^3}\left(\Sigma^C_{\alp\bet}\right)^2\right)^{
\ot} }
\label{AREA}
\ee
The universal constant $B$ here must be real, in virtue of the loop
orientation
symmetry. When the orientation is reversed $C(\theta) \ra
C(2\pi-\theta)$, the
loop integral changes sign, but its square, which enters here, stays
invariant.
Therefore, the constant in front must be real. The time reversal
tells the
same, since {\em both} viscosity $\nu$ and the energy dissipation
rate $\et$
are time-odd.  Therefore, the ratio $\frac{\et}{\nu^3}$ is time-even,
hence it
must enter $\Psi[C]$ with the real coefficient. Clearly, this
coefficient $B$
must be positive, since $ \left|\Psi[C] \right|<1$.

\section{Discussion}

So, we found  an {\em exact} solution of the loop equation in the
turbulent
limit. It remains to be seen, whether this is the most general
solution, and is
it realized in turbulent flows. Meanwhile, let us discuss its general
properties, and its implications to the large scale vorticity
distribution.

First of all, let us address the issue of the uniqueness of this
solution. Let
us take  the following Anzatz
\be
S[C] = f\left( \oint_C d \ral \oint_C d r'_{\alp} W(r-r') \right)
\ee
When substituted into the static loop equation (with the area
derivatives
computed in Appendix A), it yields the following equation for the
correlation
function $W(r)$
\bea
0=\oint_C d \ral \oint_C d r'_{\bet} \oint_C d r''_{\gam}
U_{\alp\bet\gam}(r,r',r'') \br
U_{\alp\bet\gam}(r,r',r'')=
W(r-r')\hat{V}^{\alp}_{\bet\gam}W(r'-r'') +
\mbox{permutations}\br
\hat{V}^{\alp}_{\mu\nu} = \del_{\alp\nu} \d_{\mu} -\del_{\alp\mu}
\d_{\nu}
\label{U}
\eea
The derivative $f'$ of the unknown function drops from the static
equation.

This equation should hold for arbitrary loop $C$. Using the Taylor
expansion
for the Stokes type functional \ct{Mig83}, we can argue, that the
coefficient
function $U$  must vanish up to the total derivatives. An equivalent
statement
is that the third area derivative of this functional must vanish.
Using the
loop calculus  (see Appendix A) we find the following equation
\be
0=\hat{V}^{\alp}_{\mu\nu} \hat{V'}^{\alp'}_{\mu'\nu'}
\hat{V''}^{\alp''}_{\mu''\nu''} U_{\alp\alp'\alp''}(r,r',r'')
\ee
which should hold for arbitrary $r,r',r''$.
This leads to the overcomplete system of equations for $W(r)$ in
general case.
However, for the special case $ W(r) = r^2$ which corresponds to the
square of
the tensor area
\be
\Sigma_{\alp\bet}^2 = - \oh \oint_C d \ral \oint_C d
r'_{\alp}(r-r')^2
\ee
the system is satisfied as a consequence of certain symmetry.  In
this case  we
find in the loop equation
\be
2 \oint_C d \ral \oint_C d r'_{\bet}(r-r')^2 \oint_C d r''_{\alp}
\left(
r'_{\bet} - r''_{\bet}\right) \propto  \Sigma^C_{\alp\bet}\oint_C d
\ral
\oint_C d r'_{\bet}(r-r')^2
\ee
The last integral is symmetric with respect to permutations of $ \alp
, \bet$,
whereas the first  factor $ \Sigma^C_{\alp\bet}$ is antisymmetric,
hence the
sum over $\alp\bet$ yields zero, as we already saw above.

This solution can also be used to get the probability distribution
for the
velocity circulation.  For this purpose the extra factor $\gamma$
should be
inserted in the definition of the loop average
\be
\Psi \left[C \right] = \left \langle \exp
\left(
    \frac{\i \gamma}{\nu}  \oint_C dr_{\alpha}v_{\alpha}
 \right) \right \rangle \label{eq4''}
\ee
and the Fourier transformation should be performed
\be
P_C(\Gamma) = \frac{1}{2\pi\nu}\int_{-\8}^{+\8}d\gamma \EXP{-
\frac{\i
\gamma\Gamma}{\nu} }\Psi[C]
\ee

Now, the $\gamma$ dependence of $\Psi[C]$ can be found from the
dimensional
analysis. It enters only in combination with viscosity $ \nu$, and
the other
source of viscosity, the $ \nu \d^2 v$ term, was dropped in the
Navier-Stokes
equation. So, the asymptotic solution \rf{AREA} should get the extra
factor of
$|\gamma|$ in the exponential.
\be
\Psi[C] \propto \EXP{- B\,|\gamma|\,
\left(\frac{\et}{\nu^3}\left(\Sigma^C_{\alp\bet}\right)^2\right)^{\ot
} }
\label{AREA'}
\ee
Another way to get this factor is to observe, that one could scale
$\gamma$
away from original definition, by  rescaling $\et \ra \et
|\gamma|^{-3} $,
since the velocity circulation scales as $\et^{\ot}$.

The Fourier integral yields the Lorentz distribution
\be
P_C(\Gamma) = \frac{\bar{\Gamma}}{\pi \left(\Gamma^2 +
\bar{\Gamma}^2\right)}
\\;\;
\bar{\Gamma} =
B\,\left(\et\left(\Sigma^C_{\alp\bet}\right)^2\right)^{\ot}
\ee
Let us answer an obvious  question. The Lorentz distribution is
symmetric. How
does it agree with the known asymmetry of velocity correlations, in
particular,
the Kolmogorov triple correlation?
The answer is that the Kolmogorov correlation does not imply the
asymmetry of
{\em vorticity} correlations.

Taking the tensor version of the $\frac{4}{5}$ law in arbitrary
dimension $d$
 \be
\left\langle  v_{\alpha}(0) v_{\beta}(0) v_{\gamma}(r)\right\rangle =
\frac{{\cal E }}{(d-1)(d+2)}
	\left(
	 \delta_{\alpha \gamma} r_{\beta} +
	 \delta_{\beta \gamma} r_{\alpha} -
	 \frac{2}{d}\delta_{\alpha \beta} r_{\gamma}
	\right)
\label{KOLM}
\ee
and differentiating, we find that
\be
\VEV{\val(0)\vbe(0)\omega_{\gam\lam}(r)} = 0
\ee
So, the odd vorticity correlations could, in fact, be absent, in
spite of the
asymmetry of the velocity distribution.
Besides, the area law  does not apply to small loops  which are
involved in the
definition of the vorticity correlations in terms  the loop
functional.
Moreover, we do not see any reason to expect these correlations to be
scale
invariant. From our point of view these are not the asymptotic
quantities, so
the viscous effect could be important.

With velocity correlations it is even worse than that. The infrared
divergencies may be important as well, so that the factors like $
\left(\frac{L}{r}\right)^{\delta} $ could appear in the higher
moments of
velocity distribution. The observed violations of the Kolmogorov
scaling in
these moments could be attributed to these infrared divergencies.
However,
according to our theory, the velocity circulation agrees with the
Kolmogorov
scaling and has  smooth Lorentz distribution  in the infinite system.

It would be extremely interesting to measure the velocity circulation
for large loops in real or numerical experiments, and test these
predictions.
Maybe the long tails of the Lorentz distribution  reflect the
notorious
intermittency? Note, that all the even moments are infinite for the
Lorentz
distribution, which is quite unusual, but does not contradict any
known
physical requirements.

In real world this would probably mean that the $n$-th moments of the
circulation are cut off by the finite time and finite size effects,
i.e. they
grow  as $\min\left\{\left(\et t^2\right)^{ n },\left(\et
L^4\right)^{\frac{n}{3}} \right\}$. Instead of measurements of these
moments,
the shape of the Lorentz distribution and the area dependence of its
width $
\bar{\Gamma}$ could be tested.

It was assumed in above arguments, that the loop $C$  consist of
only one
connected part. Let us now consider the more general situation, with
arbitrary
number $n$ of loops $C_1,\dots C_n$. The corresponding Anzatz would
be
\be
S_n\left[C_1,\dots C_n\right] = s_n\left(\Sigma^1,\dots
\Sigma^n\right)
\ee
where $\Sigma^i$ are tensor areas.

This function should obey the same WKB loop equations in each
variable.
Introducing the loop vorticities
\be
\omega^k_{\mu\nu}  = 2 \pbyp{s_n}{\Sigma^{k}_{\mu\nu}}
\ee
which are constant on each loop, we have to solve the following
problem. What
are the values of $ \omega^k_{\mu\nu}$ such that the single velocity
field
$\val(r) $ could produce them?

We do not see any other solutions, but the trivial one, with all
equal $
\omega^k_{\mu\nu} $ and linear velocity, as before. This would
correspond to
\be
s_n\left(\Sigma^1,\dots \Sigma^n\right)= s_1\left(\Sigma\right)\\;\;
\Sigma_{\mu\nu}=\sum_{k=1}^n \Sigma^k_{\mu\nu}= \oint_{\uplus C_k}
r_{\mu} d
r_{\nu}
\ee
The loop equation would be satisfied like before, with $ C = \uplus
C_k $. This
corresponds to the additivity of loops
\be
S_n\left[C_1,\dots C_n\right] = S_1\left[\uplus C_k\right]
\ee
Note, that such additivity is the opposite to the statistical
independence,
which would imply that
\be
S_n\left[C_1,\dots C_n\right] = \sum S_1\left[C_k\right]
\ee
The  additivity could also be understood as a statement, that any set
of $n$
loops is equivalent to a single loop for the abelian Stokes
functional. Just
connect these loops by wires, and note that the contribution of wires
cancels.
So, if the area law holds for {\em arbitrary} single loop, than it
must be
additive.

This assumption may not be true, though, as it often happens in the
WKB
approximation. There is no single asymptotic formula, but rather
collections of
different WKB regions, with quantum regions in between. In our case,
this
corresponds to the  following situation.

Take the large circular loop, for which the WKB approximation holds,
and try to
split it into two large circles. You will have to twist the loop like
the
infinity  symbol $ \8$, in which case it intersects itself. At this
point, the
WKB approximation might break, as the short distance velocity
correlation might
be important near the self-intersection point. This may explain the
paradox of
the vanishing tensor area for the $\8$ shaped loop. From the point of
view of
our area law such loop is not large at all.

This cancellation of large circulations is puzzling.
Apparently, the further analytical and numerical study of the loop
equation is
required. One could try some variational approach, by approximating
the loop by
a polygon, which reduces the $\Psi$ functional to a function of the
vertices of
the polygon. There are some other options, such as truncation of
Fourier
expansion of the loop function $ C(\theta) $. I do not expect any
easy success
here, because of singularities of the loop derivatives involved.
Still, the beauty of the loop dynamics and  its apparent reduction of
the
dimension of the turbulence problem, raises some hopes of the
analytical
advances.

The main issue, in my opinion, is the Kolmogorov scaling.  Usually,
the indexes
in the scaling solutions of the nonlinear integral equations of QFT
(so called
bootstrap equations), are  determined  from selfconsistency of the
equations,
rather than from some extra requirements. However,  the Kolmogorov
law for
triple correlation function was derived from the same \NS equation
plus the
scaling assumptions about velocity correlation functions, which is
equivalent
to the assumptions we made in our scaling Anzatz for $S[C]$. So, it
is
possible, that the dynamical value of $\kappa$, found from
selfconsistency,
would coincide with the Kolmogorov value.

However, we have to keep in mind the possibility of  the more general
phenomenon. Namely, there might be the {\em spectrum} of solutions
for the
critical index $\kappa$,  corresponding to various fixed points of
above
Hamilton-Jacobi equation in the loop space. The steady state may not
exist, if
these fixed points are all unstable. In this case, $S[C]$  would go
from one
fixed point to another.

This would be the next level of complexity, as compared to the
strange
attractors, found in dynamical systems with small number of degrees
of freedom.
Not just the trajectory in phase space, but the whole probability
distribution
functional would evolve with  with time.  When averaged over given
time
interval $T$, it would reach steady state only for the scales less
than
$T^{\kappa}$. The marginal scales distribution would slowly drift,
and larger
scales would not be in equilibrium at all. In this case the infinite
system
would never reach the steady state.

This situation is not described by above area laws, but rather
requires more
general non-steady solutions of the loop equations. In Appendices C,
D, E, F we
develop the general framework for studying such solutions. We decided
to place
these parts of our work in Appendix because this formalism is too
heavy.
However, the mathematically oriented reader might find it useful.

\section{Acknowledgments}

I am grateful to V.~Borue, I.~Goldhirsh, D.~McLaughlin, A.~Polyakov
and
V.~Yakhot for stimulating discussions .

This research  was sponsored by the Air Force Office of Scientific
Research
(AFSC) under contract F49620-91-C-0059. The United States Government
is
authorized to reproduce and distribute reprints for governmental
purposes
notwithstanding any copyright notation hereon.

\newpage

\appendix

\section{Loop Expansion}

Let us outline the method of direct iterations of the loop equation.
The full
description of the method can be found in \cite{Mig83}. The basic
idea is to
use the following representation of the loop functional
\be
\Psi[C] = 1+\sum_{n=2}^{\8} \inv{n} \left\{\oint_C dr_1^{\alp_1}
\dots  \oint_C
dr_n^{\alp_n}\right\}_{\mbox{cyclic}} W^n_{\alp_1\dots
\alp_n}\left(r_1,\dots
r_n\right)
\label{STOKES}
\ee

This representation is valid for every translation invariant
functional with
finite area derivatives (so called Stokes type functional). The
coefficient
functions $W$ can be related to these area derivatives. The
normalization
$\Psi[0]=1 $ for the shrunk loop is implied.

In general case the integration points $r_1,\dots  r_n$ in
\rf{STOKES} are
cyclicly ordered around the loop $C$. The coefficient functions can
be assumed
cyclicly symmetric without loss of generality. However,  in case of
fluid
dynamics, we are dealing with so called abelian Stokes functional.
These
functionals are characterized by completely symmetric coefficient
functions, in
which case the ordering of points can be removed, at expense of the
extra
symmetry  factor in denominator
\be
\Psi[C] = 1+\sum_{n=2}^{\8} \inv{n!} \oint_C dr_1^{\alp_1} \dots
\oint_C
dr_n^{\alp_n} W^n_{\alp_1\dots \alp_n}\left(r_1,\dots r_n\right)
\label{ABEL}
\ee
The incompressibility conditions
\be
\d_{\alp_k}W^n_{\alp_1\dots \alp_n}\left(r_1,\dots r_n\right)=0
\label{divv}
\ee
does not impose any further restrictions, because of the gauge
invariance of
the loop functionals. This invariance (nothing to do with the
symmetry of
dynamical equations!) follows from the fact, that the closed loop
integral of
any total derivative  vanishes. So, the coefficient functions are
defined
modulo such derivative terms. In effect this means, that one may
relax the
incompressibility constraints \rf{divv}, without changing the loop
functional.

To avoid confusion, let us note, that the physical incompressibility
constrains
are not neglected. They are, in fact, present in the loop equation,
where we
used the integral representation for the velocity in terms of
vorticity. Still,
the longitudinal parts of $W$ drop in the loop integrals.

The loop calculus for the abelian Stokes functional is especially
simple. The
area derivative corresponds to removal of one loop integration, and
differentiation of the corresponding coefficient function
\be
\fbyf{\Psi[C]}{\sigma_{\mu\nu}(r)} = \sum_{n=1}^{\8} \inv{n!} \oint_C
dr_1^{\alp_1} \dots  \oint_C dr_n^{\alp_n}
\hat{V}_{\mu\nu}^{\alp}W^{n+1}_{\alp,\alp_1\dots
\alp_n}\left(r,r_1,\dots
r_n\right)
\label{ABEL'}
\ee
where
\be
\hat{V}_{\mu\nu}^{\alp} \equiv
\d_{\mu}\delta_{\nu\alp}-\d_{\nu}\delta_{\mu\alp}
\ee
In the nonabelian case, there would also be the contact terms, with
$W$ at
coinciding points, coming  from the cyclic ordering \ct{Mig83}. In
abelian case
these terms are absent, since $W$ is completely symmetric.

As a next step, let us compute the local kinetic term
\be
\hat{L} \Psi[C]  \equiv \oint_C d r_{\nu}
\d_{\mu}\fbyf{\Psi[C]}{\sigma_{\mu\nu}(r)}
\ee
Using above formula for the loop derivative, we find
\be
\hat{L} \Psi[C]  = \sum_{n=1}^{\8} \inv{n!}  \oint_C dr^{\alp}\oint_C
dr_1^{\alp_1} \dots  \oint_C dr_n^{\alp_n}
\d^2W^{n+1}_{\alp,\alp_1\dots
\alp_n}\left(r,r_1,\dots r_n\right)
\label{L}
\ee
The net result is the second derivative of $W$ with respect to one
variable.
Note, that the second term in $\hat{V}_{\mu\nu}^{\alp}$ dropped, as
the total
derivative in the closed loop integral.

As for the nonlocal kinetic term, it involves the second area
derivative off
the loop, at the point $r'$, integrated over $r'$ with the
corresponding
Green's function.  Each area derivative involves the same operator
$\hat{V}$,
acting on the coefficient function. Again, the abelian Stokes
functional
simplifies the general framework of the loop calculus. The
contribution of the
wires cancels here, and the ordering does not matter, so that
\be
\frac{\delta^2\Psi[C]}
{\delta\sigma_{\mu\nu}(r)\delta\sigma_{\mu'\nu'}(r')} =
\sum_{n=0}^{\8}
\inv{n!} \oint_C dr_1^{\alp_1} \dots  \oint_C dr_n^{\alp_n}
\hat{V}_{\mu\nu}^{\alp}\hat{V'}_{\mu'\nu'}^{\alp'}W^{n+2}_{\alp,\alp'
,\alp_1\dots \alp_n}\left(r,r',r_1,\dots r_n\right)
\ee

Using these relations, we can write the steady state loop equation as
follows
\input epsf \centerline{ \epsfbox{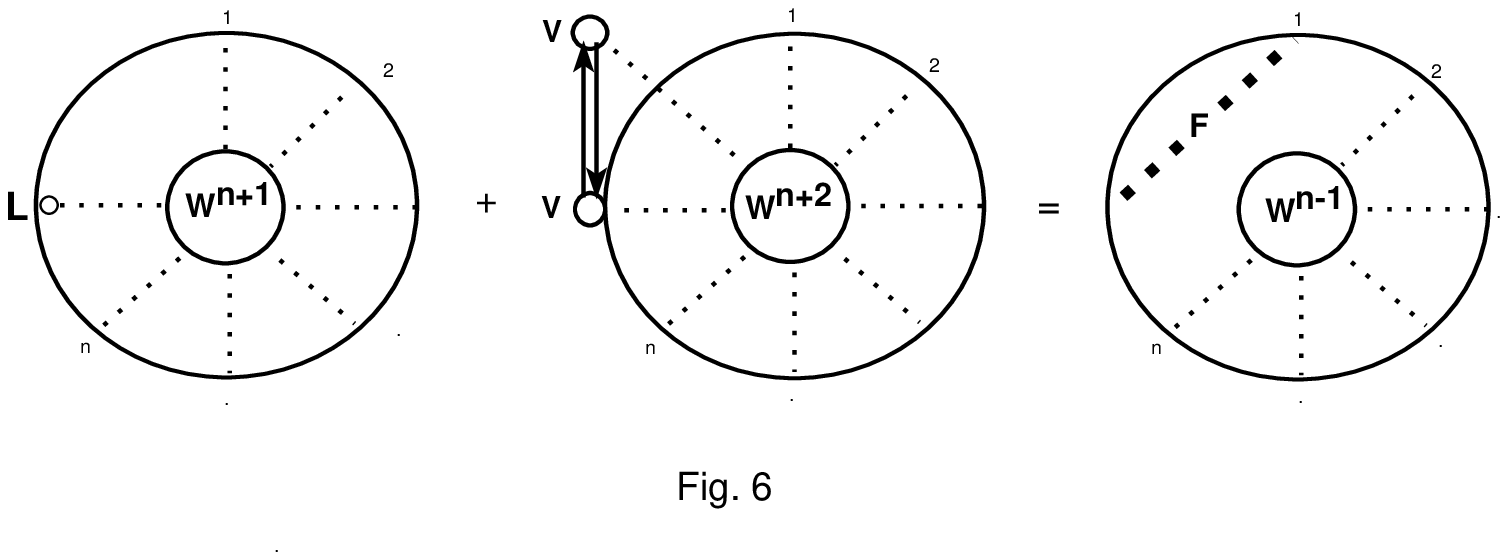}}
Here the light dotted lines symbolize the arguments $\alp_k, r_k$ of
$W$,  the
big circle denotes the loop $C$, the tiny circles stand for the loop
derivatives, and the pair of lines with the arrow denote the Green's
function.
The sum over the tensor indexes and the loop integrations over $r_k$
are
implied.

The first term is the local kinetic term, the second one is the
nonlocal
kinetic term, and the right side is the potential term in the loop
equation.
The heavy dotted line in this term stands for the correlation
function $F$ of
the random forces. Note that this term is an abelian Stokes
functional as well.

The iterations go in the potential term, starting with $\Psi[C]=1$.
In the next
approximation, only the two loop correction
$W^2_{\alp_1\alp_2}(r_1,r_2)$ is
present.  Comparing the terms, we note, that nonlocal kinetic term
reduces to
the total derivatives due to the space symmetry (in the usual terms
it would be
$ \VEV{v\omega}$ at coinciding arguments), so we are left with the
local one.

This yields the equation
\be
\nu^3 \d^2 W^2_{\alp\bet}(r-r') = F(r-r')\delta_{\alp\bet}
\ee
modulo derivative terms.  The solution is trivial in Fourier space
\be
 W^2_{\alp\bet}(r-r') = -\int \frac{d^3 k}{(2\pi)^3} \EXP{\i k
(r-r')}
\delta_{\alp\bet} \frac{\tilde{F}(k)}{\nu^3 k^2}
\ee
Note, that we did not use the transverse tensor
\be
P_{\alp\bet}(k) = \delta_{\alp\bet}- \frac{k_{\alp} k_{\bet}}{k^2}
\ee
Though such tensor is present in the physical velocity correlation,
here we may
use $\del_{\alp\bet}$ instead, as the longitudinal terms drop in the
loop
integral. This is analogous to the Feynman gauge in QED. The correct
correlator
corresponds to the Landau gauge.

The potential term generates the  four point correlation $ F \,W^2$.
which
agrees with the disconnected term in the $W^4$ on the left side
\bea
W^4_{\alp_1\alp_2\alp_3\alp_4}\left(r_1,r_2,r_3,r_4\right) \ra
W^2_{\alp_1\alp_2}\left(r_1-r_2\right)W^2_{\alp_3\alp_4}\left(r_3-r_4
\right)  +
\br
W^2_{\alp_1\alp_3}\left(r_1-r_3\right)W^2_{\alp_2\alp_4}\left(r_2-r_4
\right) +
W^2_{\alp_1\alp_4}\left(r_1-r_4\right)W^2_{\alp_2\alp_3}\left(r_2-r_3
\right)
\eea
In the same order of the loop expansion, the three point function
will show up.
The corresponding terms in kinetic part must cancel among themselves,
as the
potential term does not contribute. The local kinetic term yields the
loop
integrals of $ \d^2 W^3 $, whereas the  nonlocal one yields
$\hat{V}W^2
\,\hat{V'}W^2$, integrated over $d^3 r'$ with the Greens's function $
\frac{(r-r')}{4\pi |r-r'|^3}$. The equation has the structure
\input epsf \centerline{ \epsfbox{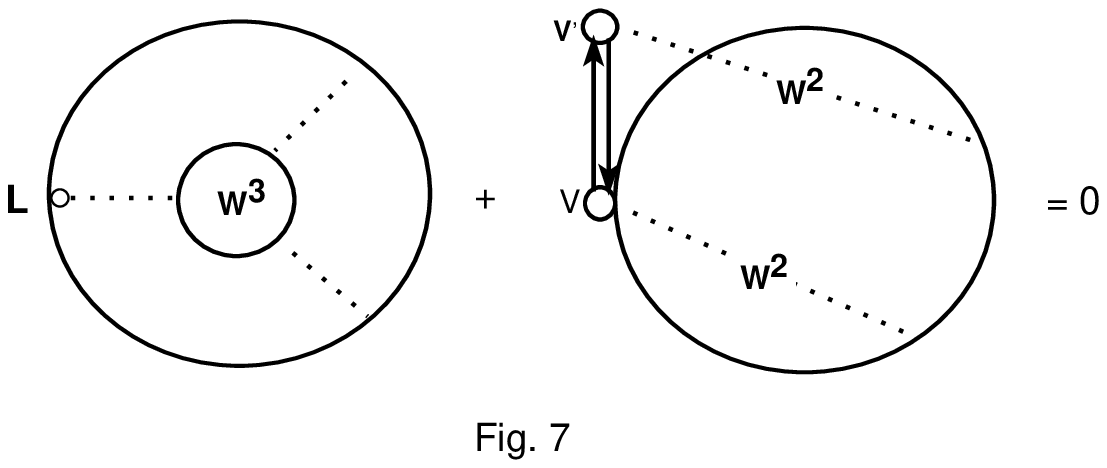}}

Now it is clear, that the solution of this equation for $W^3$ would
be the same
three point correlator, which one could obtain (much easier!) by
direct
iterations of the \NS equation.

The purpose of this painful exercise was not to give one more method
of
developing the expansion in powers of the random force. We rather
verified
that the loop equations are capable of producing the same results, as
the
ordinary chain of the equations for the correlation functions.

In above arguments, it was important, that the loop functional
belonged to the
class of the abelian Stokes functionals. Let us check that our tensor
area
Anzatz
\be
\Sigma^C_{\alp\bet}=\oint_C \ral d \rbe
\ee
belongs to the same class. Taking the square  we find
\be
\left(\Sigma^C_{\alp\bet}\right)^2 = \oint_C  d \rbe \oint_C  d
r'_{\bet} \ral
r'_{\alp} = - \oh  \oint_C  d \rbe \oint_C  d r'_{\bet} (r-r')^2
\ee
where the last transformation follows from the fact, that only the
cross term
in $ (r-r')^2$ yields nonzero after double loop integration.

Any expansion in terms  of the square of the tensor area reduces,
therefore to
the superposition of multiple loop integral of the product of
$(r_i-r_j)^2 $,
which is an example of the abelian Stokes functional.  In the limit
of large
area, this could reduce to the fractional power. An example could be,
say
\be
\Psi[C] \stackrel{?}= \EXP{B\left(1-\left(1+ \frac{\et
\left(\Sigma^C_{\alp\bet}\right)^2}{\nu^3} \right)^{\ot}\right)}
\ee
One could explicitly verify all the properties of the abelian Stokes
functional. This example is not realistic, though, as it does not
have the odd
terms of expansion. In the real world such terms are present at the
viscous
scales. According to our solution, this asymmetry disappears in
inertial range
of loops (which does not apply to velocity correlators at inertial
range, as
those correspond to shrunk loops).

\section{Matrix Model}

The Navier-Stokes equation represents a very special case of
nonlinear
PDE. There is a well known galilean invariance
\begin{equation}
v_{\alpha}(r,t) \rightarrow v_{\alpha}(r-u t,t) + u_{\alpha}
\end{equation}
which relates the magnitude of velocity field with the scales of time
and space. \footnote{At the same time it tells us that the constant
part of
velocity if frame dependent, so that it better be eliminated, if we
would like to have a smooth limit at large times. Most of notorious
large scale divergencies in turbulence are due to this unphysical
constant part.} Let us make this relation more explicit.

First, let us introduce the vorticity field
\begin{equation}
\omega_{\mu\nu} = \partial_{\mu} v_{\nu} -\partial_{\nu} v_{\mu}
\end{equation}
and rewrite the Navier-Stokes equation as follows
\begin{equation}
\dot{v}_{\alpha} = \nu \partial_{\beta} \omega_{\beta \alpha} -
v_{\beta}\omega_{\beta \alpha} - \partial_{\alpha} w \\;\;
w = p + \frac{v^2}{2}
\end{equation}

This $ w $ is the well known enthalpy density, to be found from the
incompressibility condition $ div v = 0 $, i.e.
\begin{equation}
\partial^2 w = \partial_{\alpha}v_{\beta}\omega_{\beta \alpha}
\end{equation}

As a next step, let us introduce "covariant derivative" operator
\begin{equation}
D_{\alpha} = \nu \partial_{\alpha} - \frac{1}{2}v_{\alpha}
\end{equation}
and observe that
\begin{equation}
2 \left[D_{\alpha} D_{\beta} \right] = \nu \omega_{\beta \alpha}
\end{equation}
\begin{equation}
2  D_{\beta}\left[ D_{\alpha}D_{\beta} \right] + {\it h.c.}=
\nu \partial_{\beta} \omega_{\beta \alpha} -
v_{\beta}\omega_{\beta \alpha}
\end{equation}
where $ {\it h.c.}$ stands for hermitean conjugate.

These identities allow us to write down the following dynamical
equation for the covariant derivative operator
\begin{equation}
\dot{D}_{\alpha} =  D_{\beta}\left[ D_{\alpha}D_{\beta} \right]
- D_{\alpha} W + {\it h.c.}
\end{equation}

As for the incompressibility condition, it can be written as follows
\begin{equation}
\left[D_{\alpha} D^{\dagger}_{\alpha} \right] =0
\end{equation}
The enthalpy operator $ W = \frac{w}{\nu}$ is to be determined from
this condition , or, equivalently
\begin{equation}
\left[D_{\alpha} \left[D_{\alpha} W \right] \right] =
    \left[D_{\alpha}, D_{\beta}\left[ D_{\alpha}D_{\beta}
\right]\right]
\end{equation}

We see, that the viscosity disappeared from these equations. This
paradox is resolved by extra degeneracy of this dynamics: the
antihermitean part of the $ D $ operator is conserved. Its value at
initial time is proportional to viscosity.

The operator equations are invariant with respect to the time
independent unitary transformations
\begin{equation}
D_{\alpha} \rightarrow S^{\dagger}D_{\alpha}S\\;\; S^{\dagger}S = 1
\end{equation}
and, in addition, to the  time dependent unitary
transformations with
\begin{equation}
S(t) = \exp \left( \frac{1}{2\nu} t u_\beta \left(D_{\beta} -
D_{\beta}^{\dagger} \right) \right)
\end{equation}
corresponding to the galilean transformations.

One could view the operator $ D_{\alpha} $ as the  matrix
\begin{equation}
\left\langle i | D_{\alpha} | j \right \rangle =
\int d^3r \psi_i^{\star}(r) \nu \d_{\alp} \psi_j(r)-
\frac{1}{2} \psi_i^{\star}(r)v_{\alpha}(r) \psi_j(r)
\end{equation}
where the functions $ \psi_j(r) $ are the Fourier of Tchebyshev
functions depending upon  the geometry of the problem.

The finite mode approximation would correspond to truncation of this
infinite size matrix to finite size $ N $. This is not quite the same
as leaving $ N $ terms in the mode expansion of velocity field.
The number of independent parameters here is $ O(N^2) $ rather then $
O(N)$. It is not clear whether the unitary symmetry is worth paying
such a high price in numerical simulations!

The matrix model of Navier-Stokes equation has some theoretical
beauty
and raises hopes of simple asymptotic probability distribution. The
ensemble of random hermitean matrices was recently applied to the
problem of Quantum Gravity \cite{QG}, which led to a genuine
breakthrough in the field.

Unfortunately, the model of several coupled random matrices, which is
the case here, is much more complicated then the one matrix model
studied in Quantum Gravity. The dynamics of the eigenvalues is
coupled
to the dynamics of the "angular" variables, i.e. the unitary matrices
$ S $ in above relations. We could not directly apply the technique
of
orthogonal polynomials, which was so successful in the one matrix
problem.

Another technique, which proved to be successful in QCD and Quantum
Gravity is the loop equations. This method, which we are discussing
at length in this paper, works in field theory problems with hidden
geometric
meaning. The turbulence proves to be an ideal case, much simpler then
QCD or
Quantum Gravity.

\section{The Reduced Dynamics}

Let us now try to  reproduce the dynamics of the loop field by a
simpler Anzatz
\begin{equation}
  \Psi[C] = \left \langle \exp
	\left(
	   \frac{\i}{\nu}\oint d C_{\alpha}(\theta)
P_{\alpha}(\theta)
	\right) \right \rangle \label{Reduced}
\end{equation}
The difference with original definition (\ref{eq4}) is that our new
function $ P_{\alpha}(\theta) $ depends directly on $ \theta $ rather
then through the function $ v_{\alpha}(r) $ taken at $ r_{\alpha} =
C_{\alpha}(\theta) $. This is the $ d \rightarrow 1 $ dimensional
reduction we mentioned before. From the point of view of the loop
functional there is no need to deal with field $ v(r) $ , one could
take a shortcut.

Clearly, the reduced dynamics  must be fitted to the Navier-Stokes
dynamics of original field. With the loop calculus, developed above,
we have
all the necessary tools to build this reduced dynamics.

Let us assume some unknown dynamics for the $P $ field
\begin{equation}
  \dot{P}_{\alpha}(\theta) = F_{\alpha}\left(\theta,[P] \right)
\end{equation}
and compare the time derivatives of original and reduced Anzatz. We
find in (\ref{Reduced}) instead of (\ref{Orig})
\begin{equation}
     \frac{\i}{\nu}\oint d C_{\alpha}(\theta)
	F_{\alpha}\left(\theta,[P]\right)
\end{equation}

Now we observe, that $P'$ could be replaced by the functional
derivative, acting on the exponential in (\ref{Reduced}) as follows
\begin{equation}
  \frac{\delta}{\delta C_{\alpha}(\theta)}
	\leftrightarrow  -\i\nu P'_{\alpha}(\theta)
\end{equation}
This means, that one could take the operators of the
Section 2, expressing velocity and vorticity in terms of the spike
operator, and replace the functional derivative as above.
This yields the following formula for the spike derivative
\begin{equation}
  D_{\alpha}(\theta,\epsilon) =  -\i\nu \int_{\theta}^{\theta+2
\epsilon} d \phi
	\left(
	1- \frac{\left|\theta + \epsilon - \phi \right|}{\epsilon}
	\right) P'_{\alpha}(\phi) =  -\i\nu
 \int_{-1}^{1}d \mu
	 \mbox{ sgn}(\mu)
	 P_{\alpha} \left(\theta + \epsilon (1+ \mu) \right)
	 \label{DP}
\end{equation}
This is the weighted discontinuity of the function $ P(\theta) $,
which in the naive limit $ \epsilon \rightarrow 0 $ would become the
true discontinuity. However, the function $ P(\theta) $ has in
general
the stronger singularities, then discontinuity, so that this limit
cannot be taken yet.

Anyway, we arrive at the dynamical equation for the $P$ field
\begin{equation}
  \dot{P}_{\alpha} = \nu D_{\beta} \Omega_{\beta \alpha} - V_{\beta}
\Omega_{\beta \alpha} \label{Pdot}
\end{equation}
where the operators $ V , D, \Omega $ of the Section 2 should
be regarded as the ordinary numbers, with  definition (\ref{DP}) of
$D$ in
terms of $P$.

All the  functional derivatives are gone! We needed them only to
prove equivalence of reduced dynamics to the Navier-Stokes dynamics.

The function $ P_{\alpha}(\theta) $ would become complex now,
as the right side of the reduced dynamical
equation is complex for real $ P_{\alpha}(\theta) $.

Let us discuss this puzzling issue in more detail. The origin of
imaginary units was the factor  of $ \imath $ in exponential of the
definition of the loop field. We had to insert this factor to make
the
loop field decreasing at large loops as a result of oscillations of
the phase factors. Later this factor propagated to the definition of
the $ P $ field.

Our spike derivative $ D $ is purely imaginary for real $ P $, and so
is our $ \Omega $ operator. This makes the velocity operator $ V $
real.
Therefore the $ D \Omega $ term in the
reduced equation (\ref{Pdot}) is real for real $ P $ whereas the
$ V \Omega $ term  is purely imaginary.

This does not contradict the moments equations, as we saw before. The
terms
with even/odd number of velocity fields in the loop functional are
real/imaginary, but the moments are real, as they should be. The
complex
dynamics of $ P $ simply doubles the number of independent variables.

There is one serious problem, though. Inverting the spike operator $
D_{\alpha} $ we implicitly assumed, that it was antihermitean, and
could be regularized by adding infinitesimal negative constant to $
D_{\alpha}^2 $ in denominator. This, indeed, works  perturbatively,
in
each term of expansion in time, or that in size of the loop, as we
checked.
However, beyond this expansion there would be a problem of
singularities, which arise when $ D_{\alpha}^2(\theta) $ vanishes at
some  $ \theta $.

In general, this would occur for complex $ \theta $, when the
imaginary and real part of $ D_{\alpha}^2(\theta) $ simultaneously
vanish. One could introduce the complex variable
\begin{equation}
  	e^{\imath \theta}=z\\;\;
	e^{-\imath \theta}= \frac{1}{z}\\;\;
	 \oint d \theta = \oint
	\frac{dz}{\imath z}
\end{equation}
where the contour of $z $ integration encircles the origin around the
unit circle.  Later, in course of time evolution, these contours must
be deformed, to avoid  complex roots of $ D_{\alpha}^2(\theta) $.

\section{Initial Data}
Let us study the relation between the initial data for the original
and reduced dynamics. Let us assume, that initial field is
distributed
according to some translation invariant probability distribution,
so that initial value of the loop field does not depend on the
constant part of $C(\theta)$.

One can expand translation invariant loop field in functional Fourier
transform
\begin{equation}
  \Psi[C] = \int DQ\delta^3 \left(\oint d \phi Q(\phi) \right)
	 W[Q] \exp
	\left(
	\imath \oint d \theta C_{\alpha}(\theta) Q_{\alpha}(\theta)
	\right)
\end{equation}
which can be inverted as follows
\begin{equation}
  \delta^3 \left( \oint d \phi Q(\phi)\right) W[Q] =
	\int DC\Psi[C]\exp
	\left(
	-\imath \oint d \theta C_{\alpha}(\theta) Q_{\alpha}(\theta)
	\right)
\end{equation}

Let us take a closer look at these formal transformations. The
functional measure for these integrations is defined according to the
scalar product
\begin{equation}
  (A,B) = \oint \frac{d \theta}{2 \pi} A(\theta) B(\theta)
\end{equation}
which diagonalizes in the Fourier representation
\begin{equation}
  A(\theta) = \sum_{-\infty}^{+\infty} A_n e^{\imath n \theta}
\\;\;A_{-n} = A_n^{\star}
\end{equation}
\begin{equation}
  (A,B) =  \sum_{-\infty}^{+\infty} A_n B_{-n} =
A_0 B_0 + \sum_{1}^{\infty} a'_n b'_n + a''_n b''_n\\;\;
a'_n = \sqrt{2} \Re A_n,a''_n = \sqrt{2} \Im A_n
\end{equation}

The corresponding measure is given by an infinite product of the
Euclidean measures for the imaginary and real parts of each Fourier
component
\begin{equation}
  DQ = d^3 Q_0 \prod_{1}^{\infty} d^3 q'_n d^3 q''_n
 \end{equation}
The orthogonality of Fourier transformation could now be explicitly
checked, as
\begin{eqnarray}
  \lefteqn{\int DC \exp \left( \imath \int d \theta
C_{\alpha}(\theta)
	\left(
	 A_{\alpha}(\theta) - B_{\alpha}(\theta)
	\right) \right)
	}\\ \nonumber
  &=& \int d^3 C_0 \prod_{1}^{\infty} d^3 c'_n d^3 c''_n \exp
\left( 2 \pi \imath
	\left(
	 C_0 \left(A_0-B_0 \right) +
	\sum_{1}^{\infty} c'_n\left(a'_n - b'_n \right)+
	 c''_n\left(a''_n - b''_n \right)
	\right)
\right)\\ \nonumber
&=&\delta^3\left(A_0-B_0 \right)
\prod_{1}^{\infty} \delta^3\left(a'_n - b'_n \right)
\delta^3\left(a''_n - b''_n \right)
\end{eqnarray}

Let us now check the parametric invariance
\begin{equation}
  \theta \rightarrow f(\theta)\\;\; f(2\pi) -f(0)
= 2\pi \\;\; f'(\theta) >0
\end{equation}
The functions $ C(\theta) $ and $ P(\theta) $ have
zero dimension in a sense, that only their argument transforms
\begin{equation}
  C(\theta) \rightarrow C \left( f(\theta) \right) \\;\;
 P(\theta) \rightarrow P\left( f(\theta) \right)
\end{equation}
The functions $ Q(\theta) $ and $ P'(\theta) $ in above
transformation
have dimension one
\begin{equation}
  P'(\theta) \rightarrow  f'(\theta) P'\left( f(\theta) \right)\\;\;
Q(\theta) \rightarrow  f'(\theta) Q \left( f(\theta) \right)
\end{equation}
so that the constraint on $ Q $ remains invariant
\begin{equation}
  \oint d \theta Q(\theta) = \oint df(\theta) Q\left( f(\theta)
\right)
\end{equation}

The invariance of the measure is easy to check for infinitesimal
reparametrization
\begin{equation}
  f(\theta) = \theta + \epsilon(\theta)\\;\; \epsilon(2\pi) =
\epsilon(0)
\end{equation}
which changes $C$ and $(C,C)$ as follows
\begin{equation}
  \delta C(\theta) = \epsilon(\theta) C'(\theta) \\;\;
 \delta (C,C) = \oint \frac{d \theta}{2\pi}
\epsilon(\theta) 2 C_{\alpha}(\theta) C'_{\alpha}(\theta) =
-\oint \frac{d \theta}{2\pi}\epsilon'(\theta)C_{\alpha}^2(\theta)
\end{equation}
The corresponding Jacobian reduces to
\begin{equation}
  1 - \oint d \theta \epsilon'(\theta) =1
\end{equation}
in virtue of periodicity.

This proves the parametric invariance of the functional Fourier
transformations. Using these transformations we could find the
probability distribution for the initial data of
\begin{equation}
 P_{\alpha}(\theta) = - \nu\int_{0}^{\theta} d \phi Q_{\alpha}(\phi)
\end{equation}

The simplest but still meaningful distribution of initial velocity
field is the Gaussian one, with energy concentrated in the
macroscopic
motions. The corresponding loop field reads
\begin{equation}
  \Psi_0[C] = \exp
	\left(
	 -\frac{1}{2} \oint dC_{\alpha}(\theta)
	\oint dC_{\alpha}(\theta') f\left(C(\theta)-C(\theta')\right)
	\right)
\end{equation}
where $ f(r-r') $ is the velocity correlation function
\begin{equation}
  \left \langle v_{\alpha}(r) v_{\beta}(r') \right \rangle =
\left(\delta_{\alpha \beta}- \partial_{\alpha} \partial_{\beta}
\partial_{\mu}^{-2} \right) f(r-r')
\end{equation}
The potential part drops out in the closed loop integral.

The correlation function varies at macroscopic scale, which means
that
we could expand it in Taylor series
\begin{equation}
  f(r-r') \rightarrow f_0 - f_1 (r-r')^2 + \dots \label{Taylor}
\end{equation}
The first term $ f_0 $ is proportional to initial energy density,
\begin{equation}
  \frac{1}{2} \left \langle v_{\alpha}^2 \right \rangle
=\frac{d-1}{2}
f_0
\end{equation}
and the second one is proportional to initial energy dissipation
rate
\begin{equation}
 {\cal E}_{0} = -\nu  \left \langle  v_{\alpha} \partial_{\beta}^2
v_{\alpha} \right \rangle = 2 d(d-1) \nu f_1
\end{equation}
where $ d=3 $ is dimension of space.

The constant term in (\ref{Taylor}) as well as $ r^2 + r'^2 $ terms
drop from the closed
loop integral, so we are left with the cross  term $ r r' $
\begin{equation}
  \Psi_0[C] \rightarrow  \exp
	\left(
	 - f_1 \oint dC_{\alpha}(\theta)
	\oint dC_{\alpha}(\theta')
C_{\beta}(\theta)C_{\beta}(\theta')
	\right)
\end{equation}
This is almost Gaussian distribution: it reduces to Gaussian one by
extra integration
\begin{equation}
  \Psi_0[C] \rightarrow  {\rm const }\int d^3 \omega \exp
	\left(
	 -\omega_{\alpha \beta}^2
	\right)
	\exp
	\left(
	 2\imath \sqrt{f_1}
	 \omega_{\mu\nu} \oint dC_{\mu}(\theta) C_{\nu}(\theta)
	\right)
\end{equation}
The integration here goes
over all $ \frac{d(d-1)}{2} =3 $ independent $ \alpha < \beta $
components of the antisymmetric tensor $ \omega_{\alpha \beta} $.
Note, that this is ordinary integration, not the
functional one. The physical meaning of this $ \omega $ is the random
constant vorticity at initial moment.

At fixed $ \omega $ the Gaussian functional integration over $ C $
\begin{equation}
  \int DC \exp
	\left(
	 \imath  \oint d \theta
	\left(\frac{1}{\nu}
	  C_{\beta}(\theta) P'_{\beta}(\theta)
	+2 \sqrt{f_1}
	\omega_{\alpha \beta} C'_{\alpha}(\theta)C_{\beta}(\theta)
	\right)
	\right)
\end{equation}
can be performed explicitly, it reduces to solution of the saddle
point equation
\begin{equation}
  P'_{\beta}(\theta) = 4\nu\sqrt{f_1}\omega_{\beta \alpha}
C'_{\alpha}(\theta)
\end{equation}
which is trivial for constant $ \omega $
\begin{equation}
  C_{\alpha}(\theta) =
\frac{1}{4\nu\sqrt{f_1}} \omega^{-1}_{\alpha \beta} P_{\beta}(\theta)
\end{equation}
provided the matrix $ \omega $ is invertible, which is true in
general.\footnote{The integration over $ \omega $  should be shifted
towards complex plane to avoid such degeneracy.}
We find the Gaussian
probability distribution for  $ P $ with the correlator
\begin{equation}
  \left \langle P_{\alpha}(\theta) P_{\beta}(\theta') \right \rangle
=
2\imath \nu\sqrt{f_1} \omega_{\alpha \beta} {\rm
sign}(\theta'-\theta)
\label{Corr}
\end{equation}

Note, that antisymmetry of $ \omega $ compensates that of the sign
function, so that this correlation function is symmetric, as it
should
be. However, it is antihermitean, which corresponds to purely
imaginary eigenvalues. The corresponding realization of the $ P$
functions is complex!

Let us study this phenomenon for the Fourier components.
Differentiating the last equation with respect to $ \theta $ and
Fourier transforming  we find
\begin{equation}
  \left \langle P_{\alpha,n} P_{\beta,m}  \right \rangle
= \frac{4\nu}{m} \delta_{-n m} \sqrt{f_1}\omega_{\alpha \beta}
\end{equation}

This cannot be realized at complex conjugate Fourier components $
P_{\alpha,-n} = P_{\alpha,n}^{\star} $ but we could take
$\bar{P}_{\alpha,n} \equiv P_{\alpha,-n} $ and $ P_{\alpha,n} $ as
real random  variables, with  correlation function
\begin{equation}
  \left \langle \bar{P}_{\alpha,n}P_{\beta,m} \right \rangle
= \frac{4\nu}{m}\delta_{n m}\sqrt{f_1} \omega_{\alpha \beta} \\;\;
n>0
\end{equation}
The trivial realization is
\begin{equation}
   \bar{P}_{\alpha,n} =\frac{4\nu}{n} \sqrt{f_1}\omega_{\alpha \beta}
P_{\beta,n}
\end{equation}
with $P_{\beta,n} $ being Gaussian random numbers with unit
dispersion.

As for the constant part $ P_{\alpha,0} $ of $ P_{\alpha}(\theta) $ ,
it is not defined, but it drops from equations in virtue of
translational
invariance.

\section{W-functional}

The difficulties of turbulence are hidden in the loop equation, but
they
show up, if you try to solve it numerically. The main problem is that
one cannot get rid of the cutoffs $ \epsilon, \delta \rightarrow 0 $
in the definitions of the spike derivatives. These cutoffs are
designed to pick up the singular contributions in the angular
integrals, but with finite number of modes, such as Fourier harmonics
there would be no singularities. We did not find any way to truncate
degrees of freedom in the $ P $ equation, without violating the
parametric invariance. It very well may be, that this invariance
would
be restored in the limit of large number of modes, but it looks that
there are too much ambiguity in the finite mode approximation.

After some attempts, we found the simpler version of the loop
functional,
which can be studied analytically in the turbulent region. This is
the  generating functional for the scalar products $
P_{\alpha}(\theta_1)P_{\alpha}(\theta_2) $
\begin{equation}
  W[S] = \left \langle \exp
	\left(
	 - \oint d \theta_1 \oint d \theta_2
	S(\theta_1,\theta_2)P_{\alpha}(\theta_1)P_{\alpha}(\theta_2)
	\right) \right \rangle
\label{W1}
\end{equation}
where, as before, the averaging goes over initial data for the $P $
field.

The time derivative of this W-functional
\begin{equation}
  \dot{W} = -2 \left \langle \oint d \theta_1 \oint d \theta_2
	S(\theta_1,\theta_2)P_{\alpha}(\theta_1)\dot{P}_{\alpha}(\theta_2)
	\exp
	\left(
	 - \oint d \theta_1 \oint d \theta_2
	S(\theta_1,\theta_2)P_{\alpha}(\theta_1)P_{\alpha}(\theta_2)
	\right) \right \rangle \label{W2}
\end{equation}
can be expressed in terms of functional derivatives of $W$ by
replacing
\begin{equation}
  P_{\alpha}(\phi_1)P_{\alpha}(\phi_2) \rightarrow
	\frac{\delta}{\delta S(\phi_1,\phi_2)} \label{W3}
\end{equation}
for every scalar product of $P$ fields, which arise after expansion
of
the spike derivatives (\ref{DP}), (\ref{OM}), (\ref{VOM}) in the
scalar product
\begin{equation}
  P_{\alpha}(\theta_1)\dot{P}_{\alpha}(\theta_2) =
	 \nu P_{\alpha}(\theta_1) D_{\beta}(\theta_2)
	\Omega_{\beta \alpha}(\theta_2) - P_{\alpha}(\theta_1)
	V_{\beta}(\theta_2)\Omega_{\beta \alpha}(\theta_2)
\label{PP}
\end{equation}

This equation has the structure
\begin{equation}
  \dot{W} = \oint d^2 \theta S(\theta_1,\theta_2)
	\left(
	A_2 \left[\frac{\delta}{\delta S} \right]W +
	A_3 \left[\frac{\delta}{\delta S}
\right]D^{-2}(\theta,\epsilon)W
	\right)
	\label{W4}
\end{equation}
where $A_k \left[X \right] $ stands for the $k-$ degree homogenous
functional
of
the function $ X(\theta_1,\theta_2) $.

The operator $ D^{-2} $ is also
the homogeneous functional of the negative degree $ k = -1 $. It can
be written as follows
\begin{equation}
  D^{-2}(\theta,\epsilon)W[S] =\int_{0}^{\infty}d \tau W[S + \tau U]
\end{equation}
with
\begin{equation}
  U(\theta_1,\theta_2) = \epsilon^{-2}
 \mbox{ sgn}(\theta+ \epsilon-\theta_1)
 \mbox{ sgn}(\theta+ \epsilon-\theta_2)
\end{equation}

\section{Possible Numerical Implementation}

The above general scheme is fairly abstract and complicated. Could it
lead to any practical computation method? This would depend upon the
success of the discrete approximations of the singular equations of
reduced dynamics.

The most obvious approximation would be the truncation of Fourier
expansion at some large number $ N $. With Fourier components
decreasing only as  powers of $ n $ this approximation is doubtful.
In addition, such truncation violates the parametric invariance which
looks dangerous.

It seems safer to  approximate $ P(\theta) $ by a
sum of step functions, so that it is piecewise constant. The
parametric transformations vary the lengths of  intervals of
constant $ P(\theta) $, but leave invariant these constant values.
The corresponding representation reads
\begin{equation}
  P_{\alpha}(\theta) = \sum_{l=0}^{N}
\left(p_{\alpha}(l+1)-p_{\alpha}(l)
\right)
 \Theta \left(\theta-\theta_l \right)\\;\; p(N+1) = p(1),\; p(0) = 0
\label{Thetas}
\end{equation}
It is implied that $ \theta_0 =0 < \theta_1 < \theta_2 \dots <
\theta_N < 2\pi
$.
By construction, the function $ P(\theta) $ takes value
$p(l)$ at the interval $ \theta_{l-1} < \theta < \theta_{l}
$.

We could take $ \dot{P}(\theta) $ at the middle of this interval as
approximation to $ \dot{p}(l) $.
\begin{equation}
	\dot{p}(l) \approx \dot{P}(\bar{\theta}_l)\\;\;
   \bar{\theta}_l = \frac{1}{2}\left(\theta_{l-1} + \theta_l \right)
\end{equation}
As for the time evolution of angles
$ \theta_l $ , one could differentiate (\ref{Thetas}) in time
and find
\begin{equation}
 \dot{P}_{\alpha}(\theta) =
\sum_{l=0}^{N} \left(\dot{p}_{\alpha}(l+1)-\dot{p}_{\alpha}(l)
\right)
 \Theta \left(\theta-\theta_l \right) -
\sum_{l=0}^{N} \left(p_{\alpha}(l+1)-p_{\alpha}(l) \right)
 \delta(\theta-\theta_l)\dot{\theta_l}
\end{equation}
from which one could derive the following approximation
\begin{equation}
\dot{\theta_l} \approx \frac{\left(p_{\alpha}(l)-p_{\alpha}(l+1)
\right)}{ \left(p_{\mu}(l+1)-p_{\mu}(l)\right)^2}
\int_{\bar{\theta}_l}^{\bar{\theta}_{l+1}} d \theta
\dot{P}_{\alpha}(\theta)
\end{equation}

The extra advantage of this approximation is its simplicity. All the
integrals involved in the definition of the spike derivative
(\ref{DP}) are trivial for the stepwise constant $ P(\theta) $. So,
this approximation can be in principle implemented  at the computer.
This formidable task exceeds the  scope of the present  work,
which we view as purely theoretical.

\end{document}